\documentclass[prl,aps,showpacs,twocolumn,floatfix,superscriptaddress]{revtex4}
\usepackage[dvips]{graphicx}
\usepackage{epsfig}
\usepackage{bm}

\begin{document}

\title{Pairing Reentrance Phenomenon in Heated Rotating Nuclei in the Shell Model Monte Carlo Approach}

\author{D.J. Dean}
\affiliation{Physics Division, Oak Ridge National Laboratory,
P.O. Box 2008, Oak Ridge, Tennessee 37831, USA}

\author{K. Langanke}
\affiliation{GSI Helmholtzzentrum  f\"ur Schwerionenforschung, Planckstr. 1,
D-64291 Darmstadt, Germany}
\affiliation{Institut f\"ur Kernphysik, Technische Universit\"at  Darmstadt,
Schlossgartenstr. 9,  D-64291 Darmstadt, Germany}

\author{H. Nam}
\affiliation{National Center for Computational Sciences,
Oak Ridge National Laboratory,
P.O. Box 2008, Oak Ridge, Tennessee 37831, USA}

\author{W. Nazarewicz}
\affiliation{Department of Physics and Astronomy, University of
Tennessee, Knoxville, Tennessee 37996, USA}
\affiliation{Physics Division, Oak Ridge National Laboratory,
P.O. Box 2008, Oak Ridge, Tennessee 37831, USA}
\affiliation{Institute of Theoretical Physics, University of Warsaw,
  ul. Ho\.{z}a 69, 00-681 Warsaw, Poland}

\date{\today}
\begin{abstract}
Rotational motion of  heated
$^{72}$Ge  is studied within the microscopic
Shell Model Monte Carlo approach. We investigate 
the the angular momentum alignment and nuclear pairing correlations 
associated with $J^\pi$ Cooper pairs
as a function of the
rotational frequency  and  temperature. The reentrance of
pairing correlations with temperature 
is predicted at high
rotational frequencies. It manifests itself through the  
anomalous behavior of specific heat and level density.
\end{abstract}


\maketitle

Atomic nuclei exhibit a variety of emergent
phenomena, including that of fermionic pairing  \cite{[Boh58],[Bri05]};
at low energies, the wave functions of an open-shell even-even
nucleus can be viewed as a condensate of nucleonic
Cooper pairs. Pairing plays a crucial role in our
understanding of  low-energy nuclear structure: it is
responsible for the energy gap in the spectra of even-even systems; it
gives rise to the odd-even staggering of nuclear binding energies; and
it reduces the nuclear moments of inertia
from the rigid body value \cite{[Mig59],[Bel59],[Mit62]}.  The
isovector
monopole  pairing, i.e., the coupling of like nucleons to
$J$=0 pairs, has
been identified as being the most important. With increasing temperature, nuclei undergo a distinct
transition from the pairing phase to a normal phase.  It is expected
that around $kT$=0.5-1\,MeV \cite{[Goo81a]} the static condensate vanishes.
However, as nuclei are finite  systems, this
transition is smeared out
by thermal and quantum fluctuations.  Most  theoretical
studies of the superconducting-to-normal transition have focused
principally on the relationship between pairing correlations and an
associated peak in the specific heat $C_v=dE/d(kT)$ (see, e.g.,
\cite{[Dea03],[Alh03],[Lan05],[Hou07]}). Experimental investigations of
the level densities  also indicate the possibility of the thermal
melting of the nuclear  pairing phase \cite{[Hui72],[Mel01]}.

Breaking of time-reversal symmetry by spin polarization has profound effects on nucleonic pairing. The similarity between the behavior of superconductors in external magnetic fields and the
pairing response to fast nuclear rotation  was noticed in the 1960s
\cite{[Mot60]}. 
Applying the analogy between the magnetic field  and rotational frequency
$\omega$, a sudden pairing collapse was predicted at high angular
momenta (the so-called Coriolis-antipairing effect). However, it was
realized a decade later \cite{[Bir71],[Ste72b]} that the competition
between the nuclear pairing and rotation is more intricate: the Coriolis
force acting on nucleons occupying high-$j$ orbits gradually breaks
Cooper pairs thus giving rise to an increased rotational alignment
(angular momentum polarization) and decreased collective pairing. That is, nuclei are in fact type~II superconductors in which the vortex state
can be associated with  a many-quasiparticle configuration containing broken
nucleonic pairs \cite{[Bir71]}.

Recently, there has been increased  interest in properties of
polarized (asymmetric)   Fermi systems with unusual pairing
configurations. This includes
the existence of superconductivity
in a ferromagnetically ordered phase \cite{[Pfl01],[Flo02],[Fra03],[Sch07d]}
and an  interplay between pairing, spin polarization, and
temperature in condensates having
an unequal number of spin-up and spin-down fermions
\cite{[Par06]}.
While most of the underlying theoretical discussion has been based on
schematic models or mean-field approaches,
there have been very few studies for finite systems, based on realistic interactions in
large configuration spaces, that
properly take into account quantum fluctuations \cite{[Shi89],[Hor07],[Hun08]}.

The objective of  this Letter is  to  explore nuclear
motion as a function of  temperature
and rotation by means of a realistic many-body approach. We demonstrate the presence of the phenomenon of pairing reentrance
in microscopic calculations. The effect of thermally assisted pairing 
was predicted by Kammuri in 1964 \cite{[Kam64]}. It
manifests itself as a local increase of pairing correlations in a rotating nucleus with excitation energy.  
A nice explanation of this effect was given by Moretto \cite{[Mor71]}: at $T$$\approx$0
and  large values of $\omega$,  low-energy 
states
corresponds to strongly aligned, many-quasiparticle
configurations  with strongly reduced pairing. With increasing temperature,
 less-aligned excited  configurations  with lower seniorities
enter the canonical ensemble, and this reintroduces the pair
correlations. At still higher temperatures, the superfluid-to-normal
transition takes place and pairing correlations decrease. More recent studies of thermally assisted pairing can be found in 
Refs. \cite{[Fra03],[Hun08]} based on  schematic Hamiltonians. It is worth noting that
thermally induced  pairing  is an example
of a more general  reentrance (or partial order) phenomenon manifesting itself in successive phase transitions \cite{[All81],[Rob87a],[Bal98],[For90],[Die91],[Ara05]}.

Our study is based on the Shell Model Monte Carlo (SMMC) method \cite{[Koo97]} which allows studies of nuclei at finite temperatures with the relevant degrees of freedom included. This  makes it possible to account for the thermal and quantal fluctuations which are important to describe phase transformations in finite-size systems. Within SMMC, a finite-temperature  observable is given by a  thermal average. For certain classes of residual nucleon-nucleon interactions \cite{[Lan93]},
such as the force  employed in this work,
the evaluation of observables is exact,
subject only to statistical errors related to the Monte Carlo integration.
The detailed calculations are carried out
for  a medium-mass nucleus $^{72}$Ge, which is known  to exhibit
g.s. pairing correlations and significant
quadrupole collectivity.
The details of our SMMC calculations follow closely Ref.~\cite{[Lan05]}. The single-particle (s.p.)
Fock space corresponds to the
complete $(0f1p-0g1d2s)$ shells for protons and neutrons with the s.p. energies 
(in MeV): 0 ($f_{7/2}$), 6.42 ($f_{5/2}$), 4.35 ($p_{3/2}$), 6.54 ($p_{1/2}$), 8.98 ($g_{9/2}$), 17.59 ($g_{7/2}$), 12.95 ($d_{5/2}$), 15.99 ($d_{3/2}$), and 14.64 ($s_{1/2}$).
The nucleus $^{72}$Ge
is described by
12 valence protons and 20 valence neutrons outside the closed core of $^{40}$Ca.
As we
are concerned here with a description of collective quadrupole
and pairing correlations,
we have employed a pairing+quadrupole-quadrupole Hamiltonian
with  strength parameters $G$=0.106\,MeV and $\chi$=0.0104 MeV$^{1}$fm$^2$,
which was chosen to reproduce the low-energy spectrum of
$^{64}$Ni and $^{64}$Ge. Our SMMC calculations have been performed with up to 15,840 statistical samples.
In order to generate the angular momentum polarization, we
consider the routhian $\hat{H}^\omega=\hat{H}-\omega \hat{J}_z$,
where the cranking frequency $\omega$ (in units of
MeV) enters through the cranking term.
(For early applications of SMMC to high-spin states, we refer the
reader to Refs.~\cite{[Alh96],[Dea97a]}.)

\begin{figure}[htb]
 \centerline{\includegraphics[trim=0cm 0cm 0cm 0cm,width=0.45\textwidth,clip]{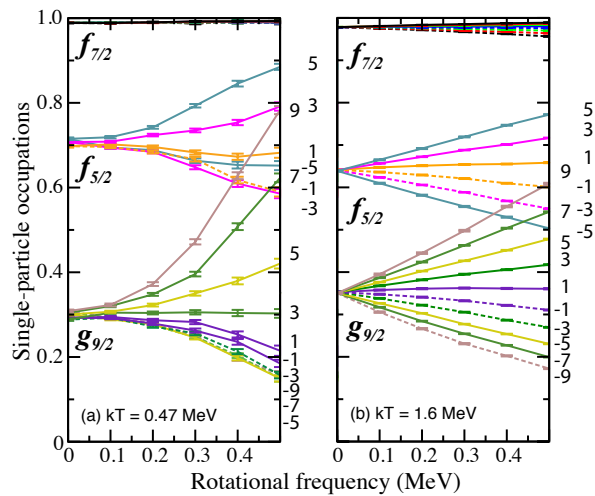}}
\caption{(Color online) Neutron
occupations of the individual $f_{7/2}$, $f_{5/2}$, and $g_{9/2}$ s.p.
orbits as a function of cranking frequency $\omega$ in the wave function
of $^{72}$Ge
for temperatures $kT$=0.47\,MeV (a) and 1.6\,MeV (b).
The angular momentum alignment in the nucleus is generated by
the gradual increase
of occupations of  positive-$m$ orbits (solid lines)
at the expense of negative-$m$ states (dashed lines).
The doubled magnetic quantum numbers, 2$m$, of individual orbits are marked.
}
\label{fig:occupations}
\end{figure}

In the absence of rotation, 
the SMMC calculations \cite{[Lan05]} gave
clear evidence for the breaking of isovector pairs at temperatures
around $kT_c \approx 0.6$ MeV in $^{72}$Ge which is reflected by a
noticeable peak in the specific heat $C_v$.
In the absence of interaction between valence nucleons, the neutrons in $^{72}$Ge would
completely occupy the $fp$ shell. However, correlations induced by
the residual interaction make it energetically favorable to scatter
neutrons across the $N$=40 shell gap which is
about 2.5\,MeV. Figure~\ref{fig:occupations} shows single-neutron
occupations
in the wave function of  $^{72}$Ge as a function of
rotational frequency at two temperatures: $kT$=0.47\,MeV (slightly
above g.s.) and 1.6\,MeV (well above $T_c$). In the g.s. configuration,
the total neutron occupation of the $gds$ shell is about 3.5, with
about 3 neutrons in the $g_{9/2}$ orbital.
As protons only occupy about half of the $fp$ shell
their excitation into the $gds$ shell is significantly smaller
(only about 0.3 protons are promoted). Upon
rotating the nucleus, the s.p. cranking term $\omega \hat{j}_z$,
representing the combined effect of the Coriolis and centrifugal force
\cite{[Szy83]}, generates angular momentum polarization by
lifting the magnetic $m$-degeneracy of s.p. states.
This is clearly seen in Fig.~\ref{fig:occupations}(a), especially for the
$g_{9/2}$, which has the largest $j$-value in our s.p. space.
For instance, at $\omega$=0.5\,MeV the occupation of
$m$=9/2 orbit  grows to $\sim$0.8  while the one of $m=$-9/2 is reduced
to $\sim$0.15 in comparison to the value of 0.3 at $\omega$=0.

The highly-asymmetric pattern of s.p. occupations seen
in Fig.~\ref{fig:occupations}(a) is due to an interplay between
the Coriolis force which tries to align individual single-particle angular
momenta along the axis of rotation, thus introducing spatial
polarization of the system, and  the symmetry-restoring
pairing force.
Indeed, in the $J$=0 pair operator
$\Delta^\dagger = \sum_{jm>0} (-1)^{j-m}a_{jm}^\dagger a_{j-m}^\dagger$,
all magnetic substates appear with the same weights.
Thus one observes
in Fig.~\ref{fig:occupations} that at low temperatures the occupations of
the various $m$ substates of the $g_{9/2}$ and $f_{5/2}$ orbitals
do not follow a simple thermal ordering. In fact, the occupations
of the substates with largest $m$ values are largest, as expected
from energy considerations, but far from one (the extreme s.p. limit).
On the other hand, the states with the lowest
negative $m$ values
have very similar reduced (but nonzero)
occupations, indicating that the residual population
 although
disfavored by the cranking term, allows the system
to gain  energy through pairing. 

\begin{figure}[htb]
 \centerline{\includegraphics[trim=0cm 0cm 0cm
 0cm,width=0.30\textwidth,clip]{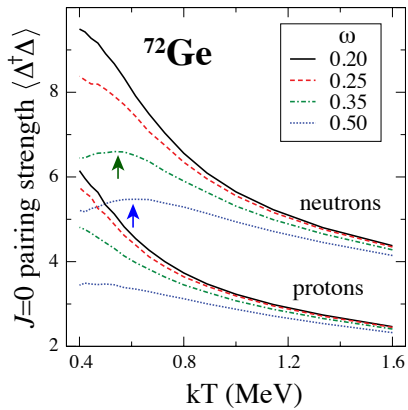}}
\caption{(Color online) Neutron (top) and proton (bottom) $J$=0 pairing
strength $\langle \Delta^\dagger \Delta \rangle$ as a function of temperature
for $^{72}$Ge at four values of $\omega$ (in MeV). At higher rotational frequencies,
neutron pairing locally {\it increases} at low temperatures
reaching a maximum (marked by arrows).
}
\label{fig:pairing-T}
\end{figure}

If the nucleus is  heated above $T_c$, 
or if the rotation is rapid,
pairing correlations
are dramatically reduced, see Fig.~\ref{fig:pairing-T}.
 However, as expected for a finite system, the superconducting-to-normal
 transition does not
occur abruptly but is smeared out
\cite{[Dea95],[Lan05]}.
Since at large temperatures pairing and quadrupole correlations are quenched,
the  occupations of s.p. states should follow the Thomas-Fermi  ordering
governed by s.p. routhians, $e^\omega_{jm}\approx e_j -\omega m$. This is precisely what is seen in
Fig.~\ref{fig:occupations}(b).

\begin{figure}[htb]
 \centerline{\includegraphics[trim=0cm 0cm 0cm 0cm,width=0.30\textwidth,clip]{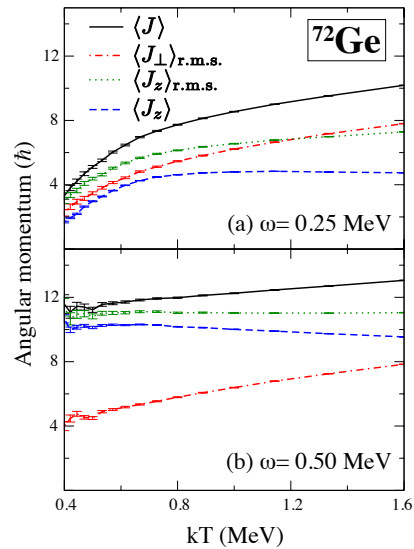}}
\caption{(Color online)
Angular momentum in $^{72}$Ge as a function of $kT$ at $\omega$=0.25\,MeV (a)
and 0.50\,MeV (b). Plotted are:  average angular momentum $\langle J \rangle$
(solid line); alignment $\langle J_z \rangle$ (dashed line);
$\langle J_\perp \rangle_{\rm r.m.s.}$ (dash-dotted line); and
$\langle J_z \rangle$ (dotted line).
 }
\label{fig:jmom}
\end{figure}

The expectation values of angular momentum components as a function
of temperature
are shown in Fig.~\ref{fig:jmom}. At low values of $\omega$,
the angular momentum alignment
$\langle J_z \rangle\equiv \langle \hat{J}_z \rangle$ is around 2.5\,$\hbar$ at the lowest temperatures considered, i.e., the
main contribution to the thermal average comes from the collective
superconducting g.s. band. At higher temperatures, the alignment increases to
$\sim$4.5\,$\hbar$, reflecting the large
contribution from the rotationally aligned
$(g_{9/2})^2$ configuration involving one broken neutron pair coupled to
$J$=8. At $\omega$=0.5\,MeV and low $T$ (Fig.~\ref{fig:jmom}b), the
lowest routhian is given by the aligned $(g_{9/2})^2$ configuration
carrying the total alignment
of  $\sim$10.5\,$\hbar$ (=2.5+8). As the nucleus is heated up, the
contributions from less aligned states become more important and
$\langle J_z \rangle$  is gradually reduced. The average angular momentum
$\langle J \rangle$, defined through the relation
$\langle \hat{J}^2 \rangle=\langle J \rangle(\langle J \rangle +1)$ steadily grows with $T$ for  low and high values of $\omega$.
At low rotational frequencies, this increase comes from  both
the parallel,
$\langle {J}_z \rangle_{\rm r.m.s.}\equiv\sqrt{\langle \hat{J}^2_z \rangle}$,
and
perpendicular angular momentum
$\langle J_\perp \rangle_{\rm r.m.s.}\equiv\sqrt{\langle \hat{J}^2-\hat{J}^2_z \rangle}$.
The situation is different at $\omega$=0.5\,MeV. Here,
$\langle J_z \rangle_{\rm r.m.s.}$$\sim$10.5\,$\hbar$
and the temperature dependence of $\langle J \rangle$ comes primarily from quantum and thermal
fluctuations in the perpendicular direction. The average moment of
inertia, defined as
$\langle J \rangle/(d\langle E\rangle/d\langle J \rangle)$,
steadily grows with $\omega$ at low temperatures, in accordance
with weakening pairing correlations. At $kT$=1.6\,MeV, the moment of inertia is
fairly constant as the  static pairing is gone.

As seen in Fig.~\ref{fig:pairing-T}, neutron pairing locally increases with
temperature at low
values of $T$ and high rotational frequencies. This is a clear signal of thermally assisted pairing.
\begin{figure}[htb]
 \centerline{\includegraphics[trim=0cm 0cm 0cm
 0cm,width=0.30\textwidth,clip]{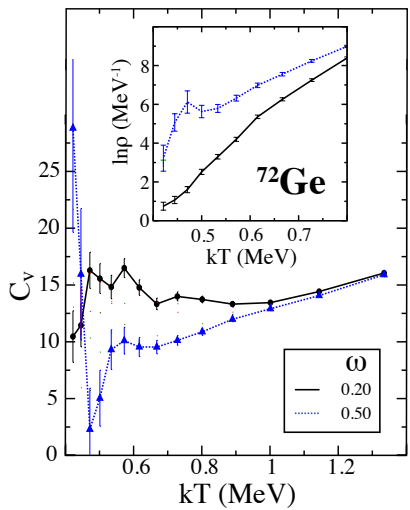}}
\caption{(Color online) Specific heat calculated in SMMC for $^{72}$Ge
as a function of temperature
at $\omega$=0.2 and 0.5 MeV. At $\omega$=0.5\,MeV,
$C_v$  shows a local dip at low temperatures;
this is a signature of the pairing reentrance. The inset shows the calculated level density. Again, at  $\omega$=0.5\,MeV, a low-temperature irregularity shows up.
}
\label{fig:cv}
\end{figure}
One of the signatures of the reentrance of the partial order is the
anomalous specific-heat behavior \cite{[Rob87a]}.
Figure~\ref{fig:cv} shows $C_v$  at different rotational frequencies.
The specific heat was calculated using the constant step $\delta\beta$=0.125 MeV$^{-1}$. Reducing 
$\delta\beta$ in the standard technique  to 0.094 or 0.0625 MeV$^{-1}$ does reproduce the behavior  in Fig.~\ref{fig:cv} and this give us confidence in the final result.
At small values of $\omega$, a typical pattern of $C_v$ is seen that is
characterized by a local maximum around $kT$=0.6 MeV
associated with breaking of pairs
\cite{[Lan05]}. This maximum is shifted to slightly lower temperatures at
higher rotational frequencies, due to  the Coriolis-antipairing effect.
At  the lowest temperatures
$C_v$ shows a sharp rise, absent at $\omega$=0 \cite{[Lan05]}. This can be attributed to a
nonzero alignment (cf. Fig.~\ref{fig:jmom})  that affects the 
nuclear moment of inertia \cite{[Alh05]}.
At $\omega$=0.5\,MeV, the specific heat exhibits a clear local dip associated
with the pairing reentrance. A pronounced low-temperature, high-frequency  irregularity is
also seen in the  level density $\rho$,  shown in the inset of Fig.~\ref{fig:cv}. The level density was obtained  according  to Refs.~\cite{[Nak97]} 
from the partition function and using
a saddle-point approximation to perform the inverse
Laplace transform. The presence of a bump in $\rho$ is not surprising: this quantity is known to strongly depend on pairing correlations; hence, can be affected by pairing reentrance \cite{[Kam64],[Hui72]}.

In summary, using the SMMC technique we explored properties
of $^{72}$Ge as a function of  temperature and  polarization. We find that
a significant contribution to the average angular momentum of a heated
nucleus comes from quantum and thermal fluctuations of the angular momentum
in the direction perpendicular to that of the average polarization.
Our  calculations demonstrate the presence of the partial order
phenomenon associated with the reappearance of pairing at high
rotational frequencies and intermediate temperatures. The signatures of such
reentrant behavior are low-temperature irregularities in the specific heat, level density, and pair transfer amplitude \cite{[Shi89]} at high rotational frequencies.

Useful comments from Nguyen Dinh Dang are gratefully appreciated. Supported by the U.S. Department of Energy under
Contract No. DE-FG02-96ER40963 (University of Tennessee). Computational resources provided by
the National Energy Research Scientific Computing Center
(Berkeley) and  the National Center for Computational
Sciencess (Oak Ridge).


\end{document}